# Universal preference for low energy core-shifted grain boundaries at the surfaces of fcc metals

**Authors:** Xiaopu Zhang*, John J. Boland*

**Affiliations:**

Centre for Research on Adaptive Nanostructures and Nanodevices (CRANN), AMBER SFI Research Centre and School of Chemistry, Trinity College Dublin, Dublin 2, Ireland

*Correspondence to: jboland@tcd.ie, xiaopuz@tcd.ie

## Abstract

Grain boundaries with [111] tilt axes are common in polycrystalline face centered cubic metals. For copper (111) films, emergent grain boundaries close to surface have tilt axes that are shifted away from [111] that are lower in energy than the corresponding truncated bulk boundaries. Geometrical analysis and atomic calculations were used to study the driving force for this same relaxation phenomenon in representative fcc elemental metals. We show that the reduction in boundary energy scales with the elimination of energetically costly boundary core facets. We find that for a wide range of misorientation angles low energy core-shifted boundaries are also favored in Al, Ni, Au and Pt and discuss the significance for electromigration and other metal properties.

## Main

The structure and stability of grain boundary defects at surfaces and interfaces control material properties such as corrosion, catalysis, and mechanical strength [1-8]. The defect or reduced binding energy relative to the bulk is responsible for electromigration-induced failure at grain boundaries in metals [9]. While strategies to stabilize grain boundaries via the dopant incorporation have been developed, we recently showed that it is possible to generate anomalously stable boundaries at the surfaces of nanoscale polycrystalline copper (111) films and bicrystals [10-12]. Experiments show these emergent grain boundaries (eGB) have tilt axes that are locally shifted away from [111] that is made possible by an out-of-plane rotation of the adjoining grains [11]. Simulations revealed a significant reduction (~20%) in the boundary energy and the formation of structures with boundary cores that lie along close packed planes. Small out-of-plane rotations lead to significant reductions in the boundary energy. The balance between boundary-energy reduction and the elastic energy cost due to grain rotation determines the depth of the low energy core-shifted boundary (CSB) [11]. For copper, CSBs can extend to many nanometers beneath the surface, approaching the dimensions of current metal interconnect technologies [13]. The formation of CSBs in the near surface region is facilitated by the increased capacity for relaxation at free surfaces. However, it remains to be established whether it is possible to engineer CSBs over larger length scales.

In this Letter, we show that this same grain boundary relaxation phenomenon occurs in other fcc metals, and we determine the degree of energy stabilization and the length scales of the CSBs in each case. To do so we developed a generalized methodology to calculate the change in boundary geometric structure during the relaxation process. We considered symmetric [111] tilt boundaries with in-plane misorientation angle $\theta$ and $(1\bar{1}0)$ boundary plane. We calculated the boundary energy as a function of the angular shift of the combined rotation axis (CRA) – due to combining the in-plane and out-of-plane rotations – along the trajectory from the original [111] tilt axis of the film (as it shifts across the $(1\bar{1}0)$ boundary plane) towards [112], which lies in the $[11\bar{1}]$ close packed plane. We then elucidated the relationship between structure change and the change in the boundary energy. Our analyses clearly show that CSBs are energetically preferred in fcc elemental metals, and we discuss the origin of this behavior and its wider importance.

We begin by constructing, in the $xyz$ coordinate system, the reference lattice from a perfect fcc lattice with its orientation $x \parallel [\bar{1}\bar{1}2]$, $y \parallel [1\bar{1}0]$ and $z \parallel [111]$, as shown in Fig.1(a). Then, the black and white lattices are generated by rotating the reference lattice by $\pm\theta/2$ along the rotation axis [111], respectively, where $\theta$ is the in-plane misorientation angle. To create the bicrystal, we choose the $xz$ plane as the boundary plane, keeping the upper half of the black lattice and similarly keeping the lower half of the white lattice (see Fig.1(b)). This process creates a symmetric tilt grain boundary with tilt axis [111], misorientation angle $\theta$, mean period vector $[\bar{1}\bar{1}2]$, and mean boundary plane $(1\bar{1}0)$. The calculation is detailed in Note S1 and selective geometrical parameters are listed in Tab. S1 [14].

To include out-of-plane rotation, we further rotate the black and white lattices by $\pm \varphi/2$ along the $y$ axis, respectively. The bicrystals before and after the out-of-plane rotation are shown as black and red in Fig. 1(c). The CRA is still in the $xz$ plane, shown in red. Note that the out-of-plane rotation axis is also in the boundary plane and parallel to the valley or ridge on the top and bottom, respectively. The resulting inclination angle $\psi$, shown in Fig. 1(c), is the angle between [111] in the bicrystal after the out-of-plane rotation and the CRA. From the two vectors in the boundary plane, the out-of-plane rotation axis and the CRA, the boundary normal is calculated. Then we calculated a new period vector of the bicrystal from the cross product of the CRA and the boundary normal. The geometry specifications for the boundaries with in-plane angle $\theta = 26.01°$ (GB26.01) with different CRAs, together with the corresponding inclination angles, are listed in Tab. S2 [14]. This method leads to much smaller unit cell sizes, making massive calculations feasible, compared with the linear combination method in our previous paper [10]. Taking the boundary with the in-plane angle $\theta = 3.89°$ and out-of-plane angle 0.79 as an example, the boundary normal is $[58\ \bar{61}\ 2]$. In our previous analysis, the period vector is $[\bar{9}\ \bar{8}17]$ and the rotation axis is [1021 1004 1013] [10]. Using the present approach based on the minimum unit cell, the two vectors are [223] and $[\bar{187}\ \bar{170}\ 238]$. The boundary area and cell volume in each cell are then about 25 times smaller.

We calculated the boundary energies in the bulk for a range of fcc metals as a function of the inclination angle of the CRA away from [111] for different misorientation angles. In all cases, the mean boundary planes are in the $(1\bar{1}0)$ plane. Boundaries were calculated with LAMMPS software, molecular statics method and the widely used embedded-atom-method (EAM) interatomic potentials for Cu, Ni, Al, Au and Pt [15-17]. To explore the effect of different potentials, we used the third-generation charge-optimized-many-body (COMB3) potential for Pt, to capture the effects of its high stacking faulting energy [18-20]. We built each repeat cell with a pair of parallel GBs of equal and opposite misorientation and used periodic boundary conditions in three directions [15,18,21]. Fully relaxed configurations were obtained by energy minimization with respect to both the atomic coordinates and the cell size along the boundary normal via a conjugate-gradient method. Structure searching with hundreds of initial configurations with different relative displacements parallel to the boundary plane was performed and only the lowest energy structures are reported in this paper [22]. Considering the 3-fold inversion symmetry of the [111] tilt axis in fcc lattices, only boundaries with misorientations from 0 to 60° are considered [23].

Results for misorientation angles between 0 and 32.20° are shown in Fig.2(a-f), with higher angle boundaries shown in Fig. S1 [14]. The misorientation angle 32.20° corresponds to the boundary [2,1] in surface notion, which has one full dislocation every two $[1\bar{1}0]$-atomic-line spacings in the Burgers vector loop. For all calculated fcc metals, the boundary energy increases monotonically with the in-plane angle from 0 to ~ 32.20°. Boundaries with [112] tilt axis ($\psi = 19.47$) are always lowest in energy, and the boundary energy decreases nearly linearly as the CRA shifts from [111] toward [112], i.e., CSBs with a [112] tilt axis are a global energy minimum regardless of the direction of the original tilt axis. The same trends are clearly seen for Pt in Fig. 2(e-f) regardless of whether the EAM or COMB3 potentials were employed. These results indicate that CSBs with [112] tilt axes in fcc metals are energetically preferred and there is no thermodynamic barrier to grain rotation.

Figure 3(a) visualizes the core structure for GB26.01 with different inclination angles and the corresponding boundary energies (mJ m$^{-2}$). The length along the CRA shown in each case is 3 times that of the vector shown in red below each panel. The boundary cores facet into segments along [112] that lie in the close packed $(11\bar{1})$ plane and evenly distributed 1/2 [110] segments that lie out of the plane, indicated by red arrows [24]. The CRA shift toward [112] reduces the boundary energy and at the same time reduces and eventually eliminates these energetically unfavorable jogs within the boundary core. For low angle boundaries, the segments along [112] prefer to dissociate into stacking fault ribbons that are a balance between decreasing the repulsive interactions between partials and minimizing the stacking fault area and energy [25]. On the other hand, each $1/2\,[110]$ segment is essentially a jog within the boundary core, shifting one stacking fault ribbon into a neighboring slip plane and in doing so it constrains the stacking fault dissociation and hence increases the total dislocation line energy.

To establish the relationship between changes in GB structure and energy, we analyzed the energy variation and the density of jogs at several inclination angles. We write, according to our calculated structures, the CRA vector or the effective boundary core direction as $n/2\,[112] + 1/2\,[110]$ with content

$$l = \sqrt{(n+1)^2 + 2n^2}/\sqrt{2}\,, \tag{1}$$

where $n \geq 1$ and positive integers are taken in our atomic calculation. For example, $n = 1$ corresponds to [111], $n = 2$ to [334], $n = 3$ to [223], $n = 5$ to [335] and $n = \infty$ to [112]. For a given value of $n$, the calculated jog line-density, that is the number of jogs per unit length, is

$$\rho_l = 1/la\,, \tag{2}$$

and the inclination angle is

$$\cos\psi = (2n+1)/\sqrt{3}l\,, \tag{3}$$

where $a$ is the lattice constant. During the out-of-plane rotation process, the initial spacing between two boundary core is changed since for any given boundary area, the dislocation line length varies as $1/\cos\psi$, which is ~ 1, since the angle $\psi$ is small. Our analysis in Fig. 3(b) shows that the boundary energy variation is nearly proportional to the jog density (the jog number per unit area) so that

$$\Delta\gamma \propto \sqrt{3}/(2n+1)\,. \tag{4}$$

The proportional coefficients (standard error) from Fig. 3(b) for GB3.89 and GB26.01 in copper are 136 (5) and 209 (9), respectively. This excellent scaling suggests that the interaction between jogs both within and between cores is small. The standard error is less than 5% of the slope for calculated boundaries with different misorientation angles in all metals, except for high angle boundaries in Pt and Al. For Pt using the COMB3 potential and Al with EAM potential, our calculated stacking fault energies (321 mJ/m$^2$ and 146 mJ/m$^2$ respectively) are high, preventing relaxation of the stresses within the cores, and hence the presence of long-range elastic interactions among jogs and neighboring cores [16,20]. As the CRA shifts in $(1\bar{1}0)$ plane from [110] toward [111] ($\psi < 0$ in Fig. 2 & S1), the $1/2\,[110]$ segments are close to each other, and the dependence shifts away from being proportional since their interaction must then be considered. As the CRA shifts from [112] toward [001] with $\psi > 19.47°$, different segments or jogs are now involved, highlighted in Fig. 3(a) using black arrows, and hence different coefficients are expected.

The linear relationship in Fig. 3(b) clearly shows that the energy reduction that drives CSB formation is achieved through removing energetically unfavorable boundary-core facets, i.e., a preference for boundary cores to lie along close packed planes. The normalized ratios of the CSB energies compared to that that of the original [111] boundaries are shown in Fig. S2 [14]. For copper, nickel and gold the energy reduction ranges from ~30% at low angle boundaries and decreases to ~10% before increasing to over 40% for GB60 twin boundaries. The corresponding ratios for aluminum are reduced, while the behavior of platinum depends on potential used, consistent with higher stacking fault energies and a reduced stabilization of dissociated boundary cores (see Fig. S2). In all cases, simulations predict an energy stabilization that is particularly significant for low angle boundaries.

To estimate the extended depth of CSBs beneath the surface, we calculated the energy difference $\Delta\gamma_{gb}$ between a bulk GB with a tilt axis [111] and its corresponding bulk boundary with its CRA shifted to [112]. We then estimated the CSB depth $h$ as

$$h = \Delta\gamma_{gb}/2cG\varphi^2 \tag{5}$$

by balancing the energetic driving force $-\Delta\gamma_{gb} \cdot h$ and the elastic energy cost $c \cdot G\varphi^2 h^2$, where $\varphi$ is the out-of-plane angle, the coefficient $c$ is related to Poisson ratio and $G$ is the shear modulus [11]. Fig. 4(a) shows a plot of the CSB depth $h$ for copper as a function of the misorientation angle with $cG = 4.35\ GPa$. A close-up view of the plot with misorientation angle greater than 15° is shown as an inset. Clearly, while the CSB depth can be many nanometers for low angles boundaries, it decreases with misorientation angle and is ultimately reduced to atomic layer thicknesses beyond ~ 32.20°. At higher angles, while there is a modest increase in the CSB depth it is still limited to several atomic layers, so it remains a surface effect.

The CSB depth for other fcc metals shown in Fig 4(b) was estimated neglecting the misorientation related elastic anisotropy. The same value $c$ was used for all metals and the value for $G$ along the metal (111) plane was obtained using elastic parameters calculated for each metal [26,27]. We see from Fig 4(b) that the variation of the CSB depth with misorientation angle is similar to that found in copper. The transition from a many nm scale tilted boundary to a surface effect still occurs at misorientation angle of ~ 30°, The notable exception is Al, where the variation ratio of the boundary energy is reduced (see Fig. S2) and hence the curve is uniformly shifted to reduced depths so that in this case the transition occurs at ~ 20°. Recognizing that the distribution of grain boundaries in polycrystalline and nanocrystalline materials is inversely correlated with the boundary energy [28], the predominance of lower angle grain boundaries suggests that the majority of the emergent boundaries at the free surfaces of fcc metals are CSBs with depths of several to many nanometers. Moreover, since these dominant low angle CSBs also exhibit the greatest level of energy stabilization (see Fig. S2), there is significant potential to mitigate against electromigration in metals by optimising grain boundary structure.

In summary, our geometrical analysis and calculations show that CSB formation is driven by the systematic removal of energetically unfavorable boundary-core facets that ultimately results in eGB with cores that lie along close packed planes. Even small levels of boundary tilting result in significant reductions in boundary energy. The energy stabilization is significant for all fcc metals, regardless of the material differences, elastic constants and stacking fault energies. Collectively, these results demonstrate that the elimination of core facets is a fundamental thermodynamic principle that drives CSB formation at the free surfaces of all fcc materials. The core-shifting phenomenon is expected to be particularly important for nanoscale metals since the CSB depth approaches the physical dimensions of these materials. In additional to electromigration, eGB structure and stability is known to impact a wide range of phenomena – grain coalescence and thin film formation, mechanical strength, electrical conductivity and catalytic activity [1,3,4,29-35] – so that additional research is needed to elucidate the role of CSBs in controlling metal properties and performance on the nanoscale.

J.J.B and X.Z. acknowledge support from Science Foundation Ireland grants (12/RC/2278 and 16/IA/4462) and thank Trinity Centre for High Performance for providing the computing. X. Z. thanks Urvesh Patil for technical support.

**Figure 1 Boundary geometry**

(a) the reference lattice adhering to the $xyz$ coordination system. (b) bicrystals and their own coordination system. (c) out-of-plane rotation $\varphi$, the combined rotation axis (CRA) $l$ and the new period vector $p$ in the new bicrystal.

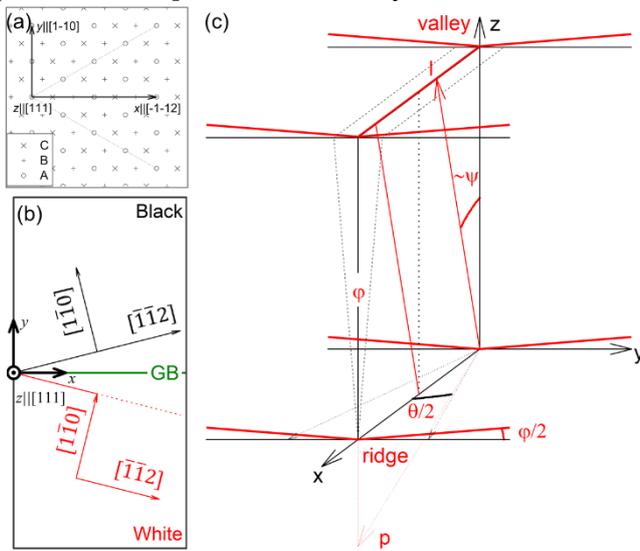

**Figure 2 inclination angle dependent GB energy for different fcc metal at different misorientation**

Inclination angle dependent GB energy for different fcc metal with misorientation angles from 0 to 32.20 for copper (a), nickel (b), aluminium (c), gold (c), platinum with EAM potential (e), and platinum with COMB3 potential (f).

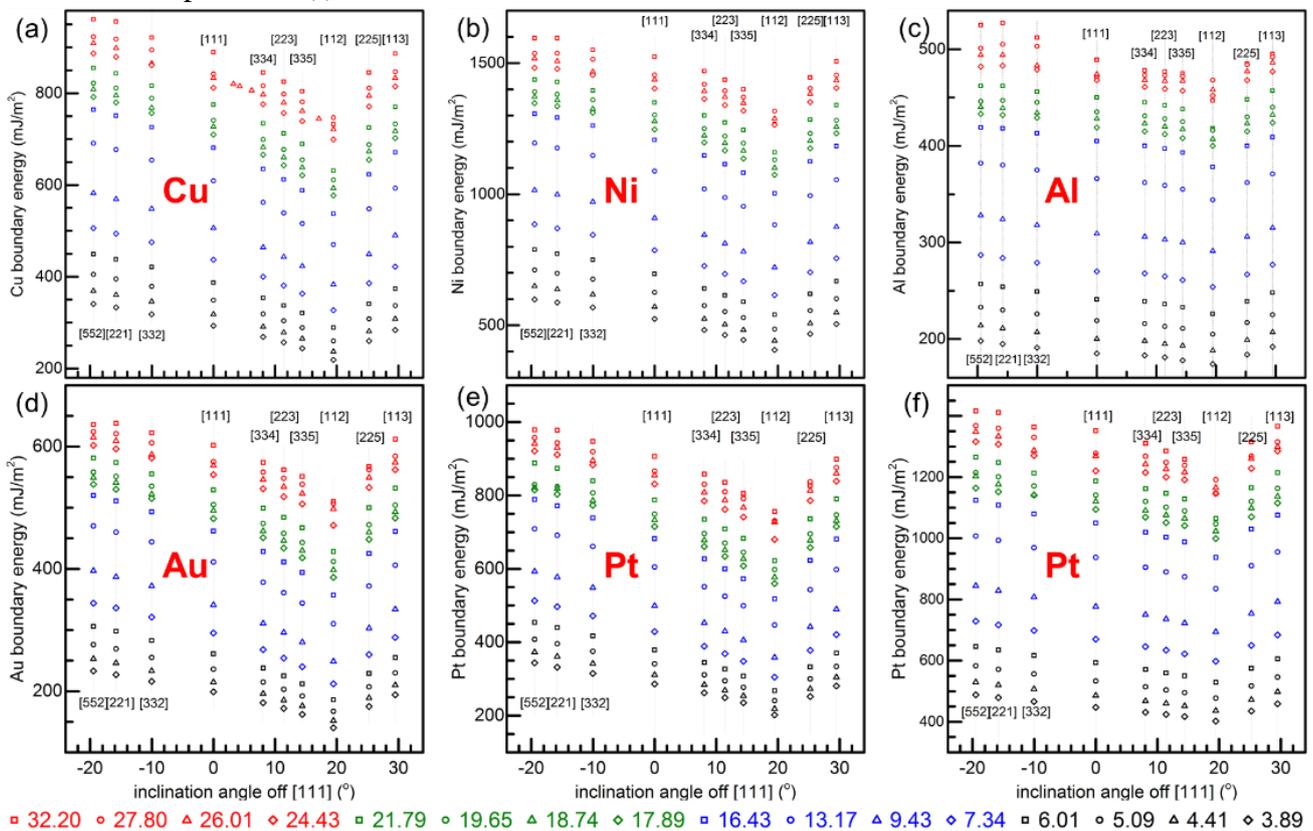

**Figure 3 Relationship between GB energy variation and the jog density.**

(a) the change of the boundary cores with in-plane angle $26.01°$ at different tilt axes in $(1\bar{1}0)$ plane, with histogram of the boundary energy (mJ m$^{-2}$). The boundaries are viewed along the direction perpendicular to the CRA and boundary normal. The length along the tilt axis is 3 times of the vector below each graph. Atoms are coloured by energy and the arrows points to jogs. (b) a plot of GB energy variation and the jog density.

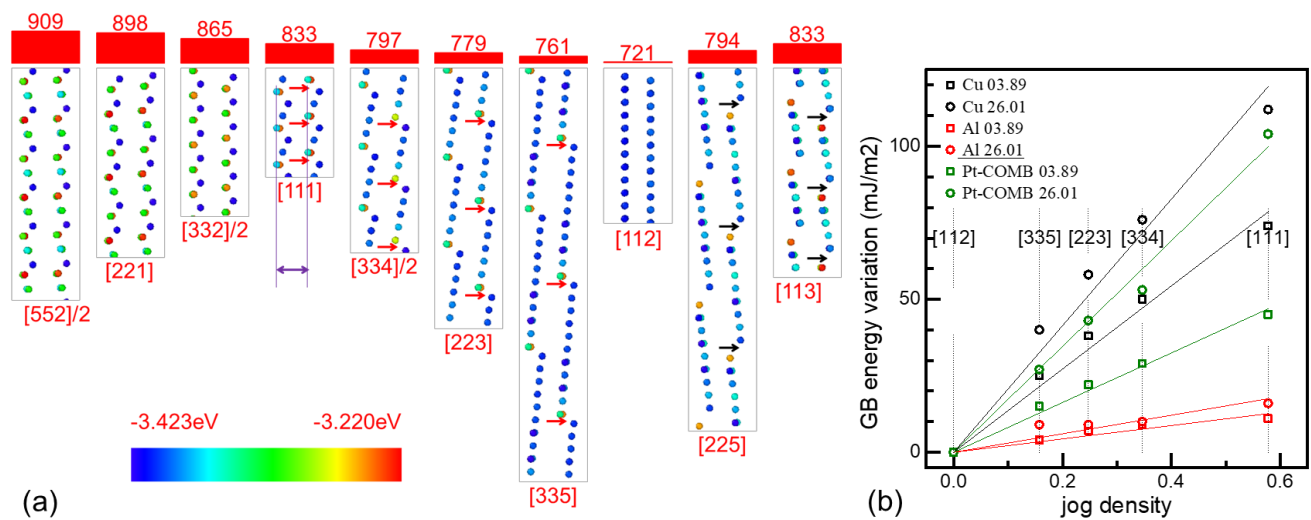

**Figure 4 misorientation dependent restructuring depth**
(a) Dots and line show the misorientation angle dependent restructuring depth. (b) the same dependence for all calculated fcc metals.

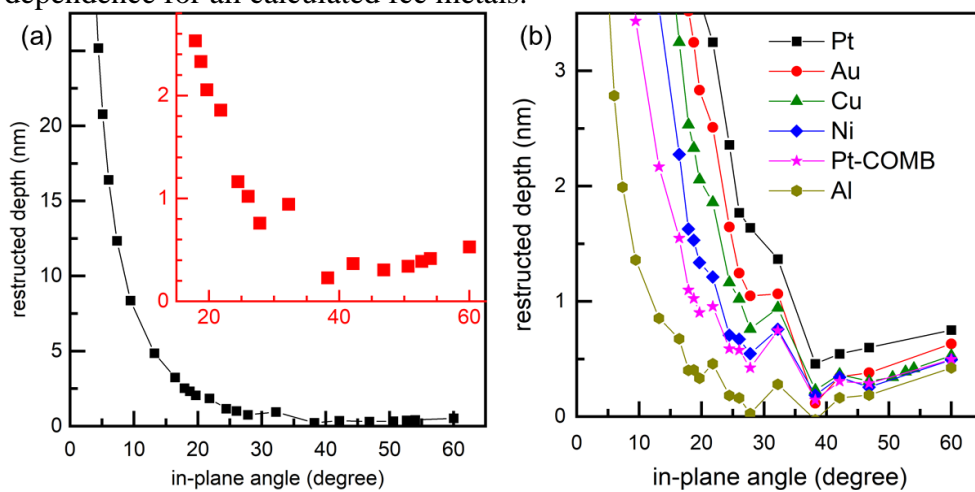